\documentclass[a4paper]{jpconf}
\usepackage{graphicx}
\usepackage{amssymb}
\usepackage{amsmath,amsfonts,amssymb,amscd,enumerate}
\usepackage{amsmath}
\usepackage{lscape}

\begin{document}
\title{Twist deformations leading to $\kappa $-Poincar\'{e}  Hopf algebra and their application to physics}

\author{Tajron Juri\'{c}$^1$, Stjepan Meljanac$^2$ 
 and Andjelo Samsarov$^{3,}$\footnote[4]{On the leave of absence from the Rudjer Bo\v skovi\'c Institute, Theoretical Physics Division, Zagreb, Croatia}}

\address{$^{1,2}$ Rudjer Bo\v skovi\'c Institute, Theoretical Physics Division, Bijeni\v cka  c.54, HR-10002 Zagreb, Croatia}
\address{$^3$ Dipartimento di Matematica e Informatica, Universita di Cagliari,
viale Merello 92, 09123 Cagliari, Italy and INFN, Sezione di Cagliari}

\ead{$^1$ tjuric@irb.hr, $^2$ meljanac@irb.hr, $^3$ samsarov@unica.it}
%\ead{samsarov@unica.it}

\begin{abstract}
  We consider two twist operators that lead to $\kappa$-Poincar\'{e} Hopf
  algebra, the first being an Abelian one and the second corresponding
  to a light-like $\kappa$-deformation of Poincar\'{e} algebra. The advantage
  of the second one is that it is expressed solely in terms of
  Poincar\'{e} generators. In contrast to this, the Abelian twist goes out
  of the boundaries of Poincar\'{e} algebra and runs into envelope of the
  general linear algebra. Some of the physical applications of these
  two different twist operators are considered. In particular, we use
  the Abelian twist to construct the statistics flip operator
  compatible with the action of deformed symmetry group. Furthermore,
  we use the light-like twist operator to define a star product and
  subsequently to formulate a free scalar field theory compatible with
  $\kappa$-Poincar\'{e} Hopf algebra and appropriate for considering the
  interacting $\phi^4 $ scalar field model on kappa-deformed space. 
\end{abstract}

\section{Introduction}
   It is known in the literature that the star product realization of
   the well known $\kappa$-deformed Minkowski spacetime algebra \cite{1,2,3,4},
   $[x_0, x_j]_* = iax_j \equiv \frac{i}{\kappa} x_j,$ for
   $j=1,...,n-1$ with remaining elements commuting \footnote{Here $a$
   is the formal parameter of deformation and $\kappa = \frac{1}{a}$
   has the mass dimension. $n$ is the number of generators of the
   spacetime algebra.}, may arise from several different types of Drinfeld's twist operators \cite{5}-\cite{12}, the Jordanian twist operator \cite{13,13a,14,15} and the Abelian twist operator \cite{16,17,18} being the most prominent ones, or at least being among those that have been mostly studied. That being said normally assumes the existence of the algebra of functions ${\mathcal{A}} [[a]] $ where the  role of the multiplication in the algebra  is being played by the twisted star product $*$ given by
  \begin{equation} \label{1}
     f * g = \mu \circ {\mathcal{F}}^{-1} (f \otimes g) = {\bar{f}}^{\alpha}(f)  {\bar{f}}_{\alpha}(g),
  \end{equation} 
  for any two functions $f$ and $g$ in the algebra and where the twist operator is being symbolically
   written  as ${\mathcal{F}} = f^{\alpha} \otimes f_{\alpha}$. 
   \footnote{${\mathcal{F}}^{-1} = {\bar{f}}^{\alpha} \otimes {\bar{f}}_{\alpha},$}
 As was pointed out in \cite{15}, in the case that this twist is   either Abelian or Jordanian, the star product (\ref{1}) leads  to the star commutation relation written above.

        On the other hand,
     when the focus is drawn toward  $\kappa$-Poincar\'{e} Hopf
     algebra \cite{1,2,3,4}, a different algebraic structure, although  closely related to
     $\kappa$-deformed Minkowski spacetime algebra 
      (see \cite{4,19} for the exact relation between the two),
 then the appearance of this structure from out of the  original Poincar\'{e} Hopf 
   algebra by means of the
     twist deformation method does not appear to be paved by as many
     proposals for the twists accomplishing this task,
  yet alone not in a format closed under the Poincar\'{e} algebra.
    Indeed, not only that there haven't been many proposals on offer, but also
 it was believed that the twist which would account for
     such a deformation does not even exist\footnote{The situation here is somewhat tricky. See the discussion in
     subsection 3.4. for a more detailed explanation.}.
     
    It is though true that some attempts have been made to get $\kappa$-Poincar\'{e} Hopf
     algebra from the Poincar\'{e} Hopf
     algebra, but neither of them finished with much of a success. The main problem being that the twist used either gives rise to
      a coalgebra (or at least the part of a coalgebra) that runs out of the Poincar\'{e}
     algebra and escapes into $igl(n)$ or the twist operator itself is
     not made  of Poincar\'{e} generators only, but in building it up
     one needs to  supplement the Poincar\'{e} generators by the elements of $igl(n)$.

% On the other hand, as far as the $\kappa$-Poincar\'{e} Hopf algebra is concerned and its appearance from the Poincar\'{e} Hopf algebra by means of the twist deformation method, then it appears that there hasn't been so %many proposals on offer.
     
 In this paper we consider two types of Drinfeld's twists that can accomplish the aforementioned deformation that leads from the undeformed Hopf algebra to the deformed one ($\kappa$-Poincar\'{e} Hopf
      algebra). The first one is the Abelian twist
\begin{equation} \label{2}
     {\mathcal{F}} = \exp \left\{  i(\lambda x_k p_k \otimes A  - (1-\lambda)A \otimes x_k p_k )  \right\},
  \end{equation} 
where $ 0 \leq \lambda \leq 1, $ is a real parameter and $A=ap_0$,
and the second one is the so called light-like twist
\begin{equation}\label{3}
   {\mathcal{F}}  = \exp \left\{ia^\alpha P^\beta\frac{\ln (1+a\cdot P)}{a\cdot P}\otimes M_{\alpha\beta}\right\},
\end{equation}
where $a^2 = 0$ (therefrom the name).
It was recently realized  that the Abelian twist (\ref{2}), despite the previous
lack of success, may though reproduce the $\kappa$-Poincar\'{e} Hopf
      algebra. However, the price paid is that one
      needs to replace the
      Hopf algebra framework with a somewhat more general algebraic
      structure, like e.g. that of the Hopf algebroid \cite{lu,xu,bohm}.
Contrary to this, when the light-like twist (\ref{3}) is concerned,
everything is smooth  there. This twist operator can  reproduce  $\kappa$-Poincar\'{e} Hopf
      algebra by staying only within the framework of the Hopf
      algebra. There is no need to go outside of this framework and to
      invoke certain more general algebraic structures. In addition to that,
      there are only Poincar\'{e} generators appearing in
      the twist (\ref{3}). 

   In what follows we briefly describe how these twists came up onto
   the surface and discuss their physical applications.
But before that, the notion of $\kappa$-Poincar\'{e} algebra in a given basis is
introduced and the relation between particular basis of $\kappa$-Poincar\'{e} and
$\kappa$-Minkowski spacetime realization, the operator ordering
prescription and the star product is explained.

\section{$\kappa$-Poincar\'{e} Hopf algebra}

We  take a brief look at  $\kappa$-Poincar\'{e}
algebra, which is one particular algebraic structure obtained by quantizing
the usual Poincar\'{e}  algebra and where the $\kappa$ is a formal
deformation parameter. Our interest shall be primarily focused on its coalgebraic
sector, particularly on the coproducts for the Poincar\'{e}
generators, that is the momentum generators
$p_{\mu} = -i\partial_{\mu},$  
\begin{eqnarray} \label{4}
{\triangle}_{\varphi} (p_{0}) & =& p_0 \otimes \mathbf{1}  + \mathbf{1} \otimes p_0, \nonumber \\ 
 {\triangle}_{\varphi} (p_{i}) & =& \varphi (A \otimes \mathbf{1} + \mathbf{1} \otimes A )
 \left[ \frac{p_i}{\varphi (A)} \otimes \mathbf{1} + e^A \otimes  \frac{p_i}{\varphi (A)} \right]
\end{eqnarray}
and the Lorentz generators $M_{\mu \nu},$
\begin{eqnarray} \label{5}
{\triangle}_{\varphi} (M_{ij}) & =& M_{ij} \otimes \mathbf{1} + \mathbf{1} \otimes M_{ij}, \nonumber  \\ 
 {\triangle}_{\varphi} (M_{i0}) & =& M_{i0} \otimes \mathbf{1} + e^A \otimes M_{i0}
 -  a p_j \frac{1}{\varphi (A)} \otimes M_{ij},
\end{eqnarray}
where $A=- ia_0 \partial_0 = ap_0. $ Here we chose $ a_{\mu} =
(a,0,...,0), ~a \sim \frac{1}{\kappa}. $

It can be noted that the coalgebra is parametrized by the function
$\varphi (A).$ It is a smooth function of $A$ and each  choice of this
function corresponds to a certain basis of  $\kappa$-Poincar\'{e}
algebra. For example, if $\varphi (A) = 1,$ then the coalgebra
(\ref{4}),(\ref{5}) corresponds to the bicrossproduct basis of
$\kappa$-Poincar\'{e}.

More direct interpretation of the function $\varphi (A)$ may be
acquired if we turn to another interesting object, which is intimately
related to  $\kappa$-Poincar\'{e} algebra: $\kappa$-Minkowski spacetime,
\begin{equation} \label{6}
  [\hat{x}_i, \hat{x}_j] = 0, \quad [\hat{x}_0, \hat{x}_j] = ia\hat{x}_j \equiv \frac{i}{\kappa}
   \hat{x}_j, \quad  \mbox{for} \quad j=1,...,n-1.  
\end{equation} 
In this context, each choice of $\varphi (A)$ corresponds to a
particular differential representation of $\kappa$-Minkowski spacetime
in terms of the generators $x_\mu, p_\nu$ of the undeformed Heisenberg
algebra
\begin{equation}\label{7}
   [x_\mu, x_\nu] = 0, \quad [p_\mu, p_\nu] = 0, \quad [p_\mu, x_\nu]
   = -i\eta_{\mu \nu},
\end{equation}
where $\eta_{\mu \nu} = diag(-1,1,...,1)$.
More explicitly, the realizations of $\kappa$-Minkowski spacetime that
correspond to the coalgebra (\ref{4}),(\ref{5}) all belong to the family
\begin{eqnarray}
 \hat{x}_i &=& x_i \varphi (A), \label{8} \\
 \hat{x}_0 &=& x_0 \psi(A) - a  x_k p_k \gamma(A),  \label{9}
\end{eqnarray}
with two other functions, $\gamma (A)$ and $\psi (A),$ not being
  independent of $\varphi (A)$, but rather being related to it through
  the consistency relation
  $\frac{\varphi^\prime}{\varphi}\psi = \gamma -1$. Let as note that there are two approaches to the tangent space of a noncommutative space whose coordinate algebra is the enveloping algebra of a Lie algebra: the Heisenberg double construction and the approach via deformed derivatives, usually defined by procedures involving orderings among noncommutative coordinates or equivalently involving realizations via formal differential operators. In \cite{skodaD} it is shown that the two different formalisms for introducing the noncommutative phase space for the Lie algebra type noncommutative spaces are mathematically equivalent. 

Furthermore, to each particular realization parametrized by the
function $\varphi (A)$ there is a uniquely defined operator ordering
prescription, so that for example one has
\begin{eqnarray}
: e^{i k_\mu  {\hat{x}}^{\mu}} :_L   &\equiv & e^{-i k_0  {\hat{x}}_{0}}
 e^{i k_i  {\hat{x}}_{i}} ~ =~ e^{-i k_0  {\hat{x}}_{0} + i k_i
 {\hat{x}}_{i} \varphi_S (ak_0) e^{ak_0}}, \nonumber \\
   : e^{i k_\mu  {\hat{x}}^{\mu}} :_R   &\equiv & e^{i k_i  {\hat{x}}_{i}} e^{-i k_0  {\hat{x}}_{0}}
  ~ =~ e^{-i k_0  {\hat{x}}_{0} + i k_i
 {\hat{x}}_{i} \varphi_S (ak_0)},  \label{10} \\
 : e^{i k_\mu  {\hat{x}}^{\mu}} :_S   &\equiv & e^{i k_\mu
 {\hat{x}}^{\mu}}, \nonumber
\end{eqnarray}
for the left, right and totaly symmetric Weyl ordering,
respectively. In the above expressions $\varphi_S (A) = \frac{A}{e^A -1}$.
Moreover, for the general ordering prescription labeled by
$:~:_\varphi$ and corresponding to an arbitrary realization $\varphi
(A)$, one can write
\begin{equation}\label{11}
   : e^{i k_\mu  {\hat{x}}^{\mu}} :_\varphi   ~ =~ e^{-i k_0  {\hat{x}}_{0} + i k_i
 {\hat{x}}_{i} \frac{\varphi_S (ak_0)}{\varphi (ak_0)} }.
\end{equation}
It is readily seen that from the later relation the left, right and Weyl totally symmetric ordering can
be recovered by taking the choices $\varphi = e^{-A}, ~ \varphi = 1, ~ \varphi =
\varphi_S, $ respectively. 
     Of interest here will be the following family of operator orderings,
\begin{equation}\label{12}
   : e^{i k_\mu  {\hat{x}}^{\mu}} :_\lambda   ~ =~ e^{-i\lambda k_0
 {\hat{x}}_{0}} e^{ i k_i {\hat{x}}_{i}} e^{-i(1-\lambda) k_0 {\hat{x}}_{0}},
\end{equation}
corresponding to the $\kappa$-Minkowski spacetime realization $\varphi (A)
= e^{-\lambda A}$. It interpolates between the right, time-symmetric
and left ordering, corresponding respectively to $\lambda = 0,
\frac{1}{2}$ and $1$. 

    The exponentials $: e^{i k_\mu  {\hat{x}}^{\mu}} :_\varphi $ may
    be viewed as the plane waves on the $\kappa$-Minkowski
    spacetime. If further they are to be considered as the elements of
    the Borel group, then the notion of coproduct encodes the group
    multiplication rule
\begin{equation}\label{13}
   : e^{i k_\mu  {\hat{x}}^{\mu}} :_\varphi : e^{i q_\mu  {\hat{x}}^{\mu}} :_\varphi   ~ =~ 
  : e^{i (k \oplus q)_\mu  {\hat{x}}^{\mu}} :_\varphi ~ =~ e^{i
  {\mathcal{D}}^{\varphi} (k,q) x},
\end{equation}
where the connection between the coproduct and the function ${\mathcal{D}}^{\varphi} (k,q)$ 
is unraveled by the relation $\triangle_{\varphi} (p_\mu) =
{\mathcal{D}}^{\varphi}_{\mu} (p \otimes 1, 1 \otimes p) $.

    Thus, to conclude, for each $\varphi (A)$ parametrizing the initial coalgebra,
    there is a unique operator ordering prescription, there is a
    unique realization of the accompanying structure ($\kappa$-Minkowski
    space) and there is a unique star product and a twist operator, as
    we shall see shortly.

A comment is needed regarding the algebra sector. In the algebra
sector, the Lorentz algebra remains undeformed in this setting, while
the commutator $[M_{\mu \nu}, p_\lambda]$ is deformed and depends on
the realization $\varphi (A)$. However, it is also possible to formulate 
the  $\kappa$-Poincar\'{e} algebra in the so called classical basis
\cite{19,20,21}. Classical here means that the algebraic
sector is undeformed and the whole deformation is contained within the
coalgebraic sector. This coalgebra is not contained within the
coalgebra parametrization (\ref{4}),(\ref{5}) and needs to be treated
separately. It looks as
\begin{eqnarray} \label{14}
 \triangle (P_{\mu}) &=& P_{\mu}\otimes Z^{-1}+\mathbf{1}\otimes
 P_{\mu}- a_{\mu} (P_{\lambda} Z)\otimes
 P^{\lambda} +\frac{a_{\mu}}{2} \square \; Z\otimes aP, \\
 \triangle (M_{\mu\nu}) &=& M_{\mu\nu}\otimes
 \mathbf{1}+\mathbf{1}\otimes M_{\mu\nu} \nonumber \\
 &+& a_{\mu}\left(-P^{\lambda}+\frac{a^{\lambda}}{2}\square\right)\,
 Z\otimes
 M_{\lambda\nu} +a_{\nu}\left(P^{\lambda}-\frac{a^{\lambda}}{2}\square\right)\,
 Z\otimes M_{\lambda\mu}, \label{15}
\end{eqnarray}
where this time $a_\mu$ is the fourvector deformation parameter and
\begin{equation} \label{16}
   Z^{-1} (P) = aP + \sqrt{1 + a^2 P^{2}}, \qquad
 \square (P) = \frac{2}{a^2}(1- \sqrt{1+ a^2 P^2}), \quad P_\mu = -i \partial_\mu.   
\end{equation}

%We now turn to antipodes for the generators of $\kappa$-Poincar\'{e} algebra.
%The antipode for translation generators $P_{\mu}$ can be
%written in a compact way as
%\begin{equation} \label{3.8}
% S(P_{\mu}) = \left( -P_{\mu} - a_{\mu}P^2 +
% \frac{1}{2}a_{\mu}(aP)\square (P) \right) Z(P),
%\end{equation}

%On the other hand, antipode for Lorentz generators has the form
%\begin{equation} \label{3.8b}
% S(M_{\mu \nu}) = -M_{\mu \nu}
%   + a_{\nu} \left ( P_{\alpha} - \frac{a_{\alpha}}{2} \square (P)
%   \right ) M_{\alpha \mu}
%  - a_{\mu} \left ( P_{\alpha} - \frac{a_{\alpha}}{2} \square (P)
%   \right ) M_{\alpha \nu}.
%\end{equation}

Similarly as before, there exists a realization of the accompanying $\kappa$-Minkowski
    space,
\begin{equation} \label{17}
 [\hat{x}_{\mu},\hat{x}_{\nu}]  = i(a_{\mu}\hat{x}_{\nu}-a_{\nu}\hat{x}_{\mu}),
\end{equation}
that corresponds to the above coalgebra and it is given by
\begin{equation} 
\hat{x}_{\mu}= x_\mu \left(aP+\sqrt{1+ a^2 P^2} \right) - (ax) P_\mu.
\label{18}
\end{equation}

\section{Twist operators and twisting}

\subsection{Abelian twist}

  Star product for the general parametrization  $\varphi (A)$ can be derived \cite{22} and the result is
   \begin{equation} \label{19}
   (f \; {\star}_{\varphi} \; g)(x)
     =  \lim_{\substack{u \rightarrow x  \\ t \rightarrow x }}
    e^{x_{j} {\partial}_{j}^{u} \left (
     \frac{ \varphi (A_{u} + A_{t}) }{\varphi (A_{u})} - 1  \right )
        + x_{j} {\partial}_{j}^{t} \left ( 
           \frac{ \varphi (A_{u} + A_{t}) }{\varphi (A_{t})} e^{A_{u}} - 1 \right )}
       f(u) g(t), 
   \end{equation}
   where $A_u=- ia_0 \partialì^u_0, \; A_t=- ia_0 \partialì^t_0. $
   It corresponds to  the coalgebra (\ref{4}),(\ref{5}) and also to the $\kappa$-Minkowski
       spacetime realization (\ref{9}), both of them being parametrized with $\varphi (A)$.
 As a next step, we recall the general definition of the star product  
\begin{equation} 
  f * g = \mu_* (f \otimes g) = \mu \circ {\mathcal{F}}^{-1} (f \otimes g)
\label{20}
\end{equation}
and the result which says that for an arbitrary operator $f$ commuting with $D \equiv x_i \partial_i$
($D$ is the dilatation operator) it holds \cite{22}
\begin{equation} \label{21}
    :e^{Df}: ~ = ~ e^{D \ln (1+f)}. 
\end{equation}
Relying on these two, it is possible to extract the twist operator from the above star product,
\begin{equation} \label{22}
   {\mathcal{F}}_\varphi ~ = ~  
    \exp \left\{ (D \otimes 1) \ln \frac{\varphi (A \otimes 1 + 1 \otimes A)}{\varphi (A \otimes 1)}
          + (1 \otimes D) (A \otimes 1 +  \ln \frac{\varphi (A \otimes 1 + 1 \otimes
    A)}{\varphi (1 \otimes A)})  \right\}.
\end{equation}
For $\varphi = e^{-\lambda A} $ this reduces to the family of Abelian twists parametrized with the real number $\lambda$,
\begin{equation} \label{23}
   {\mathcal{F}} ~ = ~  
     \exp \left\{ i(\lambda x_k p_k \otimes A - (1 - \lambda)A \otimes x_k p_k)  \right\}.
\end{equation}

\subsection{Light-like twist}

There is a straightforward, but tedious and not always feasible procedure for constructing the twist operator from a given coalgebra. The method requires introducing the coordinate generators and considering their coproducts along with the coproducts for the symmetry generators (momenta).
It also utilizes the perturbative methods which in due course give a twist operator expressed in the form of an infinite series of terms, whose summation is anything but trivial. However, the ultimate hope is that this series can be resumed and brought into a closed form. The method is the following \cite{23}:
Starting from the nontrivial coproducts $\triangle x$ and $\triangle p$, representing a given deformed coalgebra, one searches for the operator ${\mathcal{F}}$ that executes the following transformations:
\begin{equation} \label{24}
  \triangle (x_\mu ) = {\mathcal{F}} \triangle_0 (x_\mu )  {\mathcal{F}}^{-1}, \quad
   \triangle (p_\mu ) = {\mathcal{F}} \triangle_0 (p_\mu )  {\mathcal{F}}^{-1}.  
\end{equation}
Here 
\begin{equation} \label{25}
  \triangle_0 (x_\mu ) = x_\mu \otimes \mathbf{1}, \quad
   \triangle_0 (p_\mu ) = p_\mu \otimes \mathbf{1}  + \mathbf{1} \otimes p_\mu,
\end{equation}
are the undeformed coproducts for the coordinate and momentum generators. Next, one expresses 
 $\triangle x$ and $\triangle p$ as the power series in the deformation parameter $a \sim \frac{1}{\kappa}$,
\begin{equation} \label{26}
 \triangle (x_\mu) = \sum_{k=0}^{\infty} \triangle_k (x_\mu ), \quad
   \triangle (p_\mu) = \sum_{k=0}^{\infty} \triangle_k (p_\mu ), \quad  \triangle_k (x_\mu ), \triangle_k (p_\mu ) \sim a^k.
\end{equation} 
For the twist operator one may take the ansatz  ${\mathcal{F}} = e^f,$ where $f= \sum_{k=1}^{\infty} f_k$ and symbolically $ f_k \propto a^k x p^{k+1} $ are the objects that need to be found. Then by using
(\ref{24}) and comparing both sides in these expressions, order by order in the deformation parameter, one gets the set of conditions
\begin{equation}\begin{split}\label{27}
&\Delta_{1} (x_\mu) = [f_{1},\Delta_{0} (x_\mu )]\\
&\Delta_{2} (x_\mu)=[f_{2},\Delta_{0} (x_\mu)]+\frac{1}{2}[f_{1},[f_{1},\Delta_{0} (x_\mu)]]\\
&\Delta_{3} (x_\mu)=[f_{3},\Delta_{0} (x_\mu)]+\frac{1}{2}\left([f_{1},[f_{2},\Delta_{0} (x_\mu)]]+[f_{2},[f_{1},\Delta_{0} (x_\mu)]]\right)+\frac{1}{3!}[f_{1},[f_{1},[f_{1},\Delta_{0} (x_\mu)]]]\\
&...\\
&\Delta_{k} (x_\mu)=[f_{k},\Delta_{0} (x_\mu)]+...+\frac{1}{k!}[f_{1},[f_{1},...[f_{1},\Delta_{0} (x_\mu)]]]\\
&...\\
\end{split}\end{equation}
and analogously for $\Delta (p_{\mu})$. 
The interesting thing is that the above described procedure
can be carried out in a full extent, when applied to the coalgebra (\ref{14}),(\ref{15}).
Even more, in the special case when $a^2 = 0,$ the infinite sum $f= \sum_{k=1}^{\infty} f_k$
can be evaluated and written in a compact form \cite{24}. The result is
the twist operator (\ref{3}).

 \subsection{Twisting}

Taking the Abelian twist (\ref{2}) and applying it onto the primitive coproduct
$\triangle_0 (p) = p \otimes \mathbf{1} + \mathbf{1} \otimes p$ (by means of the twist deformation
 (\ref{24})), one recovers the momentum part (\ref{4}) of the coalgebra (\ref{4}),(\ref{5}).
 However, if one repeats the story and applies the Abelian twist (\ref{2}) onto the primitive coproduct of Lorentz generators, he does not get the part (\ref{5}) of the $\kappa$-Poincar\'{e} algebra!
   Nevertheless, despite this inconvenience, the Abelian twist though
   may be used to reconstruct the full coalgebra (\ref{4}),(\ref{5}),
   as well as  the appropriate antipodes, which for simplicity reasons
   were not written in this letter (see
  \cite{mali}). Though, to accomplish this, one needs to extend the algebraic framework of the Hopf algebra and  go beyond it. In this case an appeal to a broader algebraic framework, like for example that of the Hopf algebroid, appears to be necessary.
   It is also necessary to use certain tensor exchange identities \cite{rmatrix} and when twisting, one does not twist the primitive coproduct itself, but the object that is compatible with the homomorphic property of the coproduct. The twist (\ref{2}) satisfies the cocycle and counital conditions:
   \begin{eqnarray} \label{27}
      ({\mathcal{F}} \otimes \mathbf{1} )\cdot ( \triangle \otimes id){\mathcal{F}} & =&
       (\mathbf{1} \otimes {\mathcal{F}}))\cdot ( id \otimes \triangle ){\mathcal{F}}, \nonumber \\
        \mu \circ (\epsilon \otimes id){\mathcal{F}} &= &\mathbf{1} = \mu \circ (id \otimes \epsilon){\mathcal{F}}.
  \end{eqnarray}

The twist (\ref{3}) for the light-like $\kappa$-deformation ($a^2 = 0)$
also complies with the above conditions of cocyclicity and normalization. However, the advantage of this twist is that unlike the Abelian one, it is expressed in terms of Poincar\'{e} generators only (it doesn't contain the dilatation operator). Also, it is written in a covariant form. And finally, via twist deformation, it gives rise \cite{Juric:2015hda} to the $\kappa$-Poincar\'{e} Hopf algebra written in a classical basis,
\begin{equation}\begin{split} \label{28}
\Delta^{\mathcal{F}} (M_{\mu\nu})&=\mathcal{F}\Delta_0 (M_{\mu\nu})\mathcal{F}^{-1}= \Delta_0 (M_{\mu\nu})+
(\delta^\alpha_\mu a_\nu-\delta^\alpha_\nu a_\mu)\left(
P^\beta+\frac{1}{2}a^\beta P^2
\right)Z\otimes M_{\alpha\beta},  \\
\Delta^{\mathcal{F}} (P_{\mu})&=\mathcal{F}\Delta_0 (P_{\mu})\mathcal{F}^{-1}= \Delta_0 (P_\mu) +\left[
P_\mu a^\alpha - a_\mu
\left( P^\alpha + \frac{1}{2}a^\alpha P^2 \right)Z
\right]\otimes P_\alpha,\\
S^{\mathcal{F}}(M_{\mu\nu})&=\chi S(M_{\mu\nu}) \chi^{-1}= -M_{\mu\nu} +(-a_\mu \delta^\beta_\nu+a_\nu \delta^\beta_\mu) \left(P^\alpha + \frac{1}{2}a^\alpha P^2 \right)M_{\alpha\beta},\\
S^{\mathcal{F}}(P_{\mu})&=\chi S(P_{\mu}) \chi^{-1}=\left[-P_\mu -a_\mu \left(P_\alpha + \frac{1}{2}a_\alpha P^2 \right) P^\alpha \right]Z,\\
\epsilon^{\mathcal{F}}(M_{\mu\nu})&=0, \quad \epsilon^{\mathcal{F}}(P_{\mu})=0,
\end{split}\end{equation}
where $Z=\frac{1}{1+aP}$ and $\chi^{-1}=\mu \circ \left[(S\otimes
  1)\mathcal{F}^{-1}\right]$, with $S$ and $\epsilon$ denoting the
antiopode and counit, respectively and with the superscript ${\mathcal{F}}$ on this quantities referring to their deformed counterparts.
Therefore, the twist (\ref{3})  reproduces the coalgebra (\ref{14}),(\ref{15}) for the special case when $a^2 =0$,
that is it reproduces a deformed Poincar\'{e} Hopf algebra corresponding to a light-like deformation.
The important point to be emphasized here is that to derive
$\kappa$-Poincar\'{e} Hopf algebra by using the twist (\ref{3}), one
doesn't need to go beyond the Hopf algebra setting, since everything can be done within this framework.

%%%%%%%%%%%%%%%%%%%%%%%

\subsection{Where we stand?}

Few comments are following in order to put the picture 
presented so far into a  right perspective. If one considers the original
$\kappa$-Minkowski space, where the deformation fourvector is oriented
in the time direction, $ a_\mu =(a,0,0,0)$, together with the corresponding $\kappa $-Poincar\'{e}  Hopf algebra
which describes the symmetries of that space,
then it can be shown that there is no Drinfeld twist for this Hopf
algebra. Namely, in that case it can be shown that  no classical $r$-matrix which is
expressible only in terms of  Poincar\'{e} generators can be
constructed that satisfies CYBE
(classical Yang Baxter equation). Instead, these classical
$r$-matrices satisfy MYBE (modified Yang Baxter equation) with the right
hand side proportional to $a^2$. From here then follows that there is
no Drinfeld twist for $\kappa $-Poincar\'{e}  Hopf algebra
if $ a_\mu $ is oriented in the time direction only. If we however permit $ a_\mu $ to
point in an arbitrary direction, then for the light-like case, $a^2 =0,$
CYBE will be satisfied and there exists a Drinfeld twist for that case.
This twist is given by the formula (\ref{3}). It has to be said that
this twist is already   known for some time \cite{13} as the extended
Jordanian twist, although in a form somewhat less transparent, written  as a product of two other twists.

The interesting point is that even in the case of standard $\kappa $-Poincar\'{e}
there exists a twist operator which, besides giving rise to a correct star
product, is also by its properties very
close to the Drinfeld twist, though in a sense of the Hopf algebroid.
This twist is given by the formula (\ref{2}).

Another interesting point is that  the Hopf algebroid
framework, with the operative techniques demonstrated between the formulas
(\ref{24}) and (\ref{27}), not only  can
lead to twist operators that satisfy the cocycle condition in the
sense of the Hopf algebroid, but it can also give rise to true
Drinfeld twists,  the twist (\ref{3}) being the explicit example
of that.
%%%%%%%%%%%%%%%%%%%%%%%%

\section{Applications}
 \subsection{Twisted statistics}

We use this example to illustrate the application of the Abelian twist (\ref{2}).
Suppose first that we have a standard situation meaning that there are no  effects
induced by the noncommutativity ($a=0$).
Then assume that we have a certain symmetry algebra ${\mathcal{G}}$
(e.g. Poincar\'{e}) under which the system under
consideration is invariant. Then the action of this symmetry algebra on
the Hilbert space (here denoted by ${\mathcal{H}}$) of physical states is
realized in some representation of ${\mathcal{G}}$. Therefore, if $|\phi \rangle$
is some state in the Hilbert space ${\mathcal{H}}$, and $\Lambda$ is
some element in the symmetry algebra, then the action of ${\mathcal{G}}$ on
${\mathcal{H}}$
is realized in some representation $D$ of ${\mathcal{G}}$, $|\phi \rangle
\longrightarrow D(\Lambda) ~ |\phi \rangle $. If we further want to
extend the action of this symmetry group from one-particle to
two-particle (and generally many-particle states), we need to use the
notion of the coproduct, so that the action of ${\mathcal{G}}$ on the two-particle
states is performed as,
\begin{equation}
 |\phi \rangle \otimes |\psi \rangle \longrightarrow 
 (D \otimes D) \Delta_0(\Lambda)  |\phi \rangle \otimes |\psi \rangle,
\end{equation}
where for $a=0$ the coproduct $\Delta_0 $ is defined as $
 \Delta_0 ~ :~ \Lambda \longrightarrow \Lambda \otimes \mathbf{1} +
 \mathbf{1} \otimes \Lambda. $
Furthermore, this coproduct has to be compatible with the
multiplication $\mu$ (usual pointwise multiplication) in the algebra ${\mathcal{H}}$ of physical states.
This is achieved by the requirement
\begin{equation}
 \mu \, \left ((D \otimes D) \Delta_0(\Lambda) |\phi \rangle \otimes |\psi \rangle \right )
= D(\Lambda) \, \mu  (|\phi \rangle \otimes |\psi \rangle).
\end{equation}

%***************************************************************
%This statistics flip operator serves to project out irreducible
%totally symmetric/antysymmetric subspaces with respect to a
%deformed/twisted symmetry group (algebra). (Crucial is to project out
%irreducible subspaces with respect to deformed symmetry group with
%vectors/tensors of a certain symmetry type. Process of
%(anty)symmetrization here may be different)

%**********************************************

When considering quantum mechanics and its premise of
indistinguishable particles, the quantum statistics is usually
implemented by restricting to subspaces which are composed of either totally symmetric
or totally antisymmetric states. This is achieved by considering the
so called statistics flip operator $\tau_0$, which
on an element $ |\phi \rangle \otimes |\psi \rangle $ from $ {\mathcal{H}} \otimes  {\mathcal{H}}$ has the action
\begin{equation} 
 {\tau}_{0}(|\phi \rangle \otimes |\psi \rangle) = |\psi \rangle \otimes |\phi \rangle,
\end{equation}
 and by using the projector  $\frac{1}{2} ( 1 \pm \tau_0 )$, with
 which help the
symmetrization or antisymmetrization on a two particle Hilbert space is carried out.
If we further want that these subspaces, built out by projectors, remain
invariant under the action of the symmetry algebra (meaning that the
process of (anti)symmetrization is invariant under the action of the
symmetry algebra), we need to require that this $\tau_0$ must commute
with $\Delta_0 $ coproduct, through which the symmetry is implemented,
\begin{equation}
[{\Delta}_{0} (\Lambda), {\tau}_{0}] = 0.
\end{equation}
 Physically this means that the process of symmetrization or
antisymmetrization is frame independent. The statistics thus remains
invariant under the action of the symmetry algebra ${\mathcal{G}}$.

%This condition can be described in the terms of a commutative diagram given by
%\begin{diagram} [width=6em]
%f \otimes g & \rTo^{\Delta_0} & (D \otimes D) \Delta_0(\Lambda) f \otimes g\\
%\dTo^{\mu} & & \dTo_{\mu} \\
%\mu (f \otimes g) & \rTo^{\Delta_0} & D(\Lambda) \mu ( f \otimes g ) \\
%\end{diagram}

In the noncommutative case, when $ a = \frac{1}{\kappa} \neq 0$, the
deformed coproduct
 $~ \Delta_\varphi = {{\mathcal{F}}_{\varphi}}^{-1}  {\Delta}_{0}
 {\mathcal{F}}_{\varphi}, ~$ where  ${\mathcal{F}}_\varphi$ is the twist element in (\ref{22}),
 will no more be compatible with
the multiplication $\mu.$ Therefore $\mu$ must be replaced with
the new multiplication $\mu_*, $ so that the compatibility condition
be satisfied,
\begin{equation}
 \mu_* \, \left ((D \otimes D) \Delta_\varphi (\Lambda) |\phi \rangle \otimes |\psi \rangle \right )
= D(\Lambda) \, \mu_*  (|\phi \rangle \otimes |\psi \rangle).
\end{equation}

 What about subspaces projected with $\frac{1}{2} ( 1 \pm \tau_0 )$?
The subspaces obtained in this way will no more be invariant under the
action of the symmetry algebra. To correct this, one introduces a new
statistics flip operator by means of the Abelian twist deformation,
 $~\tau_\varphi = {\mathcal{F}}_\varphi^{-1} \tau_0
 {\mathcal{F}}_\varphi.~$ Unlike $\tau_0$, this new $\tau_\varphi$
 will now commute with $~ \Delta_\varphi ~$ and consequently may
 serve to project out subspaces irreducibly invariant under the twisted symmetry
 algebra. The role of the projector in this case is played by the
 operator $\frac{1}{2} ( 1 \pm \tau_\varphi ).$ It gives rise to irreducible tensors
 % irreducible with respect to the considered symmetry algebra
  of certain symmetry type.

Recall now the Abelian twist   (\ref{22}). One can use it to find the
twisted statistics flip operator corresponding to the general class of
realizations (\ref{8}),(\ref{9}) of $\kappa$-Minkowski space, 
\begin{equation} \label{29}
\tau_\varphi = e^{(D \otimes A - A \otimes D)}\tau_0,
\end{equation}
and to find the corresponding universal $R$-matrix. The result for  
$\tau_\varphi$ may be used to derive the oscillator algebra of
creation and annihilation operators for the scalar field on the
noncommutative (NC) $\kappa$-deformed spacetime and to investigate the
quantum particle statistics on it. This was illustrated in the example
of the scalar field probing the background of the NC BTZ \cite{Cumesa}.
  Deformed oscillator algebras of the similar or slightly different type on $\kappa$-deformed spaces  are also considered in \cite{24.4,24.5,young,24.6,25,25.4,miao,25.5}. Moreover, the twist (\ref{23}) or its variant for $\lambda = \frac{1}{2}$ was used to construct $U(1)$ gauge theory in \cite{lma1,lma2}.
Note that as $A$ is directly proportional to the deformation parameter $a$,
the twisted flip operator $\tau_\varphi$ goes over smoothly to the untwisted flip operator $\tau_0$ as the deformation parameter $a \rightarrow 0$.

\subsection{Scalar field propagation in the ${\phi}^4 $ $\kappa$-Minkowski model}

We use this example to illustrate the application of the light-like twist (\ref{3}).
Here we consider the properties of the massive scalar field
propagation within the $\phi^4$  $\kappa$-Minkowski model based on the
light-like twist  (\ref{3}). Before writing the appropriate action, it
is worthy to note \cite{lorentzinv}  that the star
product induced by the twist (\ref{3}), which we denote by $\star,$ has the property 
\begin{eqnarray} 
\int d^n x \;\phi^{\dagger} {\star} \psi = \int d^n x \;\phi^{*} \cdot \psi\,.
\label{30}
\end{eqnarray}
That is, under the integration sign the star product ${\star}$ between two
functions may simply be dropped out and replaced by the pointwise multiplication.
As a consequence, the free part (the first two terms) of the NC action
%**************+
%The star product obtained from this light-like  twist leads to a free
%scalar field theory on $\kappa$-Minkowski space that is equivalent to
%a commutative one on a usual Minkowski space. 
%**************
\begin{eqnarray} 
\label{31}
S_n[\phi] & = & \int d^n x ~ (\partial_{\mu}\phi)^{\dagger} {\star} (\partial^{\mu}\phi) + 
  m^2 \int d^n x ~ \phi^{\dagger} {\star} \phi    
\nonumber \\    
& + &\frac{\lambda}{4}  
\int d^n x~\frac{1}{2}(\phi^{\dagger}{\star}\phi^{\dagger}{\star} \phi {\star} \phi 
    +\phi^{\dagger} {\star} \phi {\star} \phi^{\dagger} {\star} \phi)\,,
\end{eqnarray}
which is built by using the same star product, will be completely
reduced to a usual commutative free scalar field action. The only
nontrivial contributions in the action (\ref{31}) will come from the
interaction terms. As for the interaction part of the action, a
comment is in order. Actually, there are altogether six
interaction terms that could in principle appear in the action
(\ref{31}), corresponding to six possible permutations between the
fields  $\phi^{\dagger}$ and $\phi$. However, due to the properties of
the star product, they reduce to only two mutually nonequivalent
combinations and these two are that appearing in (\ref{31}).  

After expanding the star product, the effective action emerges,
\begin{eqnarray} 
  S_n[\phi] 
 & = & \int d^n x ~ \Big[(\partial_{\mu}\phi^{*}) (\partial^{\mu}\phi) +
 m^2\;\phi^{*}\phi + \frac{\lambda}{4}{(\phi^{*}\phi)}^2\Big] 
 \label{32}\\
 & + & 
 i\frac{\lambda}{4}\int d^n x \bigg[ a_{\mu} ~ 
 x^{\mu} \Big({\phi^{*}}^2(\partial_{\nu} \phi) {\partial}^{\nu} \phi -  
 \phi^{2}(\partial_{\nu} {\phi}^{*}) {\partial}^{\nu} {\phi}^{*}\Big)  
\nonumber \\
 & + & 
 a_{\nu} ~ x^{\mu} \Big(\phi^{2}(\partial_{\mu} \phi^{*}) {\partial}^{\nu} \phi^{*} -  
 {\phi^{*}}^2(\partial_{\mu} {\phi}) {\partial}^{\nu} {\phi}\Big)  
%  \nonumber \\
 % &  + &       
 +   \frac{1}{2}  a_{\nu} x^{\mu}~ \phi^{*}\phi \Big( 
    (\partial_{\mu} {\phi}^{*}) {\partial}^{\nu} \phi
     -(\partial_{\mu} \phi) {\partial}^{\nu} {\phi}^{*}\Big)  \bigg ],
 %   + {\mathcal{O}}(a^2)  
    \nonumber 
\end{eqnarray}
which can be treated within the standard framework of perturbative
quantum field theory. This means that we may find the Feynman rules
resulting from this effective action, by calculating $2$-point and
$4$-point Green's functions. In a momentum space they look as
\begin{equation} 
G\equiv G(k_1,k_2) = \frac{i}{k_1^2+m^2}\delta^{(n)}(k_1-k_2),
 \label{33}
\end{equation}
and for
the vertex function in the momentum space,
\begin{equation}
\tilde\Gamma (k_1,k_2,k_3,k_4;a)=i\frac{\delta^4 S[\tilde \phi]}{\delta\tilde\phi(k_1) 
\delta\tilde\phi(k_2)\delta\tilde\phi^*(k_3) \delta \tilde \phi^*(k_4)} \,,
\label{34}
\end{equation} 
 amounting to the following expression:
%\begin{figure}
%\begin{center}
%\includegraphics[width=40mm]{vertex.eps}
%\end{center}
%\caption{Scalar 4-field vertex}
%\label{fig:vertex}
%\end{figure}
\begin{eqnarray}
\nonumber
 \tilde\Gamma(k_1,k_2,k_3,k_4;a) &= &i(2\pi)^n\frac{\lambda}{2}\,a_{\nu} 
\Bigg[\frac{a_\nu}{a^2}+ 
\frac{1}{4}\bigg(k_{4\mu} k_{3\nu}+k_{3\mu}k_{4\nu}-2\delta_{\mu\nu}k_{4\rho}k_{3\rho}\\
\nonumber
&&\hspace{-2.5cm}
+\frac{1}{2}(k_{2\mu}k_{4\nu}-k_{4\mu}k_{2\nu}+k_{2\mu}k_{3\nu}-k_{3\mu}k_{2\nu})\bigg)
\partial_\mu^{k_1}\\
\nonumber
&& \hspace{-3cm}
+ \frac 14 \bigg( k_{4\mu} k_{3 \nu} + k_{3\mu} k_{4\nu} 
- 2 \delta_{\mu\nu} k_{4\rho} k_{3\rho} \\
\nonumber
&& \hspace{-2.5cm}
+ \frac 12 ( k_{1\mu}k_{4\nu} - k_{4\mu}k_{1\nu} + k_{1\mu}k_{3\nu} - k_{3\mu}k_{1\nu} )
\bigg) \partial_\mu^{k_2}\\
%%
%\nonumber
&& \hspace{-3cm}
+ \frac 14 \bigg(
k_{1\mu}k_{2\nu} + k_{2\mu}k_{1\nu} - 2 \delta_{\mu\nu} k_{1\rho}k_{2\rho}
\bigg) ( \partial_\mu^{k_3} + \partial_\mu^{k_4} )
\Bigg] \delta^{(n)} (k_1 + k_2 - k_3 - k_4),
\label{35}
\end{eqnarray}
where we denote $\partial_\mu^k = \frac{\partial}{\partial k_\mu}$, 
and all four momenta $k_i$ are flowing into the vertex.
The coupling $\lambda$ has to be dimensionally regularized (see \cite{phito4} for details).
Thus, while the free propagator remains unchanged, the vertex
function gets heavily modified. 

  There is one more thing that one has to take into account when
  considering the scalar field theory with a deformed underlying symmetry.
 Namely, due to a deformation of the symmetry which in this case is not the standard
  Poincar\'{e}, but rather  its quantum deformed counterpart disguised  in the
  form of $\kappa$-Poincar\'{e} symmetry, the coalgebraic part of the
  symmetry will suffer from drastic changes. Since the coalgebra,
  particularly the coproduct for the momentum generator, is
  closely related to the momentum addition rule, the connection being
  disclosed by
\begin{equation}
  \Delta^{\mathcal{F}}p_{\mu} \equiv \mathcal{D}_{\mu}(p\otimes 1,1\otimes
  p), \quad \quad \mathcal{D}_{\mu}(p, k) = p \oplus k,
\end{equation} 
 it is clear that the
  symmetry deformation will give rise to a reinterpretation of the 
  energy-momentum conservation, with the actual form of the energy-momentum conservation being dictated by the
form of coproduct for the momentum generators. We want to implement this conclusion 
  into our formalism.

Before doing  that, recall the coproduct from (\ref{28}) which is
relevant in this case, 
 \begin{equation}
\Delta^{\mathcal{F}} (P_{\mu})=\mathcal{F}\Delta_0 ( P_{\mu}) \mathcal{F}^{-1}= \Delta_0 (P_\mu) +\left[
P_\mu a^\alpha - a_\mu
\left( P^\alpha + \frac{1}{2}a^\alpha P^2 \right)Z
\right]\otimes P_\alpha,  \nonumber
\end{equation}
with $Z=\frac{1}{1+aP}$.
The corresponding expansion up to the first order in $a$ is
\begin{equation}\label{36}
\mathcal{D}_{\mu}(p,q)= p \oplus q = p_{\mu}(1+aq)+q_{\mu}-a_{\mu}\frac{pq}{1+ap}-\frac{1}{2}a_{\mu}(aq)\frac{p^2}{1+ap}.
\end{equation}

Going back to the vertex function (\ref{35}), we consistently modify
it in accordance with Eq.(\ref{36}),
 so as to keep the track with the change in the coalgebra sector of
the underlying symmetry. 
The modification includes the intervention in the argument of the
$\delta$-function in (\ref{35}),
resulting with the
Feynman rule which respects the
$\kappa$-deformed momentum addition/subtraction rule:
\begin{eqnarray} \label{37}
\nonumber
 \tilde\Gamma(k_1,k_2,k_3,k_4;a) &= &i(2\pi)^n\frac{\lambda}{2}\,a_{\nu} 
\Bigg[\frac{a_\nu}{a^2}+ 
\frac{1}{4}\bigg(k_{4\mu} k_{3\nu}+k_{3\mu}k_{4\nu}-2\delta_{\mu\nu}k_{4\rho}k_{3\rho}\\
\nonumber
&&\hspace{-2.5cm}
+\frac{1}{2}(k_{2\mu}k_{4\nu}-k_{4\mu}k_{2\nu}+k_{2\mu}k_{3\nu}-k_{3\mu}k_{2\nu})\bigg)
\partial_\mu^{k_1}\\
\nonumber
&& \hspace{-3cm}
+ \frac 14 \bigg( k_{4\mu} k_{3 \nu} + k_{3\mu} k_{4\nu} 
- 2 \delta_{\mu\nu} k_{4\rho} k_{3\rho} \\
\nonumber
&& \hspace{-2.5cm}
+ \frac 12 ( k_{1\mu}k_{4\nu} - k_{4\mu}k_{1\nu} + k_{1\mu}k_{3\nu} - k_{3\mu}k_{1\nu} )
\bigg) \partial_\mu^{k_2}\\
\nonumber
&& \hspace{-3cm}
+ \frac 14 \bigg(
k_{1\mu}k_{2\nu} + k_{2\mu}k_{1\nu} - 2 \delta_{\mu\nu} k_{1\rho}k_{2\rho}
\bigg) ( \partial_\mu^{k_3} + \partial_\mu^{k_4} )\Bigg]\\
&& \hspace{-2.5cm}\times \bigg[
\delta^{(n)} ((k_1 \oplus k_2) \ominus (k_3 \oplus k_4))
+\delta^{(n)} ((k_1 \oplus k_2) \ominus (k_4 \oplus k_3))
\bigg]\,.
\end{eqnarray}
For the more detailed explanation, see \cite{phito4}.

 In order to investigate the  properties of the massive scalar field propagation, it is most convenient
 to consider the connected $2$-point Green's function
 $G_{(c,2)}^{a}$. The important piece of information for calculating
 the connected $2$-point Green's function comes from the tad pole  diagram  contribution to the self-energy of the scalar particle, which is here calculated by using the Feynman rules (\ref{33}),(\ref{37}),
\begin{eqnarray}
\hspace{-.5cm}
{\Pi}^{a}_2=
%{\Pi}^{0,0}_2+{\Pi}^{a\ne 0,0}_2=
%+{\Pi}^{0,\xi\ne 0}_2+{\Pi}^{a\ne 0,\xi\ne 0}_2=  
%T^{0,0}_2+T^{a,0}_2+T^{0,\xi}_2=
\int \frac{d^n\ell}{(2\pi)^n} \;
\tilde\Gamma(k_1,\ell,\ell,k_4;a,\mu)\,\frac{i}{\ell^2+m^2}\,.
\label{38}
\end{eqnarray}
After regularizing $(\ref{38})$ and isolating the divergences, one gets
\begin{equation}
{\Pi}^{a}_2=\frac{\lambda m^2}{32\pi^2} \left[(1+3ak)\bigg(   
\frac{2}{\epsilon}+\psi(2)+{\rm log}\frac{4\pi\mu^2}{m^2}\bigg)
-\frac{9}{4}ak
% -(4-\epsilon)128\pi^2\frac{\xi^2 {\mu}^{4-\epsilon}}{m^8}
\right]\,.
\label{39}
\end{equation}

Since $\Pi^a_2$ is UV divergent, the Green's function $G_{(c,2)}^{a}$ will also diverge,
\begin{eqnarray}
G_{(c,2)}^{a}(k_1,k_4)&\propto&%(2\pi)^n\delta^{(n)}(k_1-k_4)
\bigg[\frac{i}{k_1^2+m^2}+
\frac{i}{k_1^2+m^2}{\Pi}^{a}_2\frac{i}{k_1^2+m^2} +...\bigg]\,,
\label{40}  \nonumber
\end{eqnarray}
\begin{eqnarray}
G_{(c,2)}^{a}(k_1,k_4)
 &{\longrightarrow}&\, 
 (2\pi)^n\delta^{(n)}(k_1 - k_4)\bigg[\frac{i}{k_1^2+m^2-{\Pi}^{a}_2}\bigg]\,.
\label{41}
\end{eqnarray}
To remove the infinities from $G_{(c,2)}^{a}$, that is to renormalize it, one adds a mass counterterm to the Lagrangian. As a consequence, the resulting Green's function ${\tilde G}_{(c,2)}^{a},$ with the contribution from the mass counterterm included, is finite. More explicitly,
\begin{eqnarray}
{\tilde G}_{(c,2)}^{a}(k_1,k_4)&=&
\Bigg[G_{(c,2)}^{a}(k_1,k_4)+
G_{(c,2)}^{a}(k_1,k_4)(-\delta m^2)G_{(c,2)}^{a}(k_1,k_4) +...\Bigg]
%_{\xi\to 0}
\nonumber\\
 &{\longrightarrow}& (2\pi)^n\delta^{(n)}(k_1-k_4)
\bigg[\frac{i}{k_1^2+m^2+\delta m^2  
 -{\Pi}^{a}_2}\bigg]\,,
\label{42}
\end{eqnarray} 
where ${\tilde G}_{(c,2)}^{a}$ denotes the Green's function 
which includes the contribution from the mass counter term.
It is easily seen that it can be rendered finite if 
\begin{equation}
\delta m^2  = \frac{\lambda m^2}{32\pi^2}    
\Bigg[(1+3ak)\frac{2}{\epsilon} 
+f\bigg(\frac{4-\epsilon}{2},\frac{\mu^2}{m^2},ak\bigg)
\Bigg].
\label{43}
\end{equation}
Therefore, in the appropriate mass counterterm the UV divergent part
is scaled with the  particle statistics factor, which depends linearly
on the noncommutativity parameter $a$ and the ingoing
momentum. Further discussion can be found in \cite{phito4}.

%Renormalization of mass by adding a mass counterterm

\section*{Acknowledgments}
This article is prepared as the contribution to the XXIII ISQS
 conference on Integrable Systems and Quantum Symmetries, Prague,
 Czech Republic. 
 A.S. acknowledges support by the European Commission and the Croatian Ministry of Science, Education and Sports through grant project financed under the Marie Curie FP7-PEOPLE-2011-COFUND, project NEWFELPRO for providing the necessary funds for attending the conference and presenting this work.
  The work by T.J. and S.M. has been fully supported by Croatian Science Foundation under the project (IP-2014-09-9582).

\smallskip

\medskip

%\item Horowitz G T and Maldacena J 2004 The black hole final state {\it J. High Energy Phys.}   %JHEP02(2004)008

%\item Kurata M 1982 {\it Numerical Analysis for Semiconductor Devices} (Lexington, MA: Heath)

%\begin{figure}[h]
%\begin{minipage}{14pc}
%\includegraphics[width=14pc]{kappaphi4-v7.ps}
%\caption{\label{label}Figure caption for first of two sided figures.}
%\end{minipage}\hspace{2pc}%
%\begin{minipage}{14pc}
%\includegraphics[width=14pc]{kappaphi4-v7.ps}
%\caption{\label{label}Figure caption for second of two sided figures.}
%\end{minipage} 
%\end{figure}

%\begin{figure}
%\begin{center}
%\includegraphics{kappa.eps}
%\end{center}
%\caption{\label{label}Figure caption}
%\end{figure}

\end{document}